# Enhancing Secrecy Energy Efficiency in RIS-Aided Aerial Mobile Edge Computing Networks: A Deep Reinforcement Learning Approach

Aly Sabri Abdalla and Vuk Marojevic
Department of Electrical and Computer Engineering, Mississippi State University, MS 39762, USA
Email: asa298@msstate.edu, vuk.marojevic@msstate.edu

***Abstract*—This paper studies the problem of securing task offloading transmissions from ground users against ground eavesdropping threats. Our study introduces a reconfigurable intelligent surface (RIS)-aided unmanned aerial vehicle (UAV)-mobile edge computing (MEC) scheme to enhance the secure task offloading while minimizing the energy consumption of the UAV subject to task completion constraints. Leveraging a data-driven approach, we propose a comprehensive optimization strategy that jointly optimizes the aerial MEC (AMEC)'s trajectory, task offloading partitioning, UE transmission scheduling, and RIS phase shifts. Our objective centers on optimizing the secrecy energy efficiency (SEE) of UE task offloading transmissions while preserving the AMEC's energy resources and meeting the task completion time requirements. Numerical results show that the proposed solution can effectively safeguard legitimate task offloading transmissions while preserving AMEC energy.**

## I. Introduction

In the advancing landscape of mobile edge computing (MEC) architectures, unmanned aerial vehicles (UAVs) have been proposed to enhance the Quality of Service (QoS) in scenarios demanding substantial computing power coupled with minimal latency. Incorporating UAVs within MEC enables leveraging their deployment flexibility. Notably, UAVs can swiftly extend communication coverage while ensuring user connectivity, especially where ground-based networks cannot provide sufficient computing resources or in cases of high user demand. Nonetheless, UAV-assisted MEC presents several challenges, attributed to inherent limitations of UAVs' power reserves and computational prowess, resulting in significant operational costs and escalated power and computing demands [1]. UAV-aided MEC operations, particularly within dense urban deployments, are frequently disrupted by obstacles such as buildings, which significantly impair line-of-sight (LoS) air-to-ground (A2G) connectivity.

Research on reconfigurable intelligent surfaces (RIS) presents an important advancement toward next-generation wireless communication networks. Comprising a multitude of cost-effective planar metasurfaces consisting of an array of passive antenna elements, the RIS phase shifters are electronically manipulated for precise control over radio frequency (RF) propagation and design of intelligent radio environments. Smart radio environments, empowered by RIS technology, play a crucial role in optimizing data transmission and processing, effectively mitigating disruptions in connectivity and QoS degradations inherent in uncontrolled wireless channels. RIS-enhanced smart radio environments offer promising avenues for spectrum sharing and optimizing security of wireless communication networks.

A RIS can be strategically deployed for improving the RF propagation in UAV-aided MEC scenarios to facilitate reliable, resilient, and high data rate communication links, particularly in environments characterized by a high probability of non-LoS (NLoS) conditions, such as dense urban areas with increased blockages. By dynamically adjusting the phase and amplitude of electromagnetic waves, the RIS can create focused beams or redirect signals, effectively establishing LoS links even in challenging urban environments. Recent research has studied RIS-assisted techniques within UAV-aided MEC. Reference [2] proposes a secure multi-layer MEC system employing a UAV with a RIS as an aerial edge server. The paper assists task offloading from multiple ground users to a base station (BS) in the presence of an active eavesdropper by iteratively optimizing RIS phase shifts, UAV deployment, power, and computing resource allocation subject to power constraints while maximizing the total number of secure computing tasks among all users. The authors of [3] leverage RIS in UAV-MEC networks and propose a double deep Q-network (DDQN) to minimize UAV energy consumption by optimizing task offloading decisions, UAV computing resource allocation, communication resource allocation, and RIS phase shifts. The work presented in [4] develops a joint multi-UAV path planning/transmission scheduling algorithm that maximizes the number of computing tasks and performance of UAVs and minimizes the total energy consumption of UAVs through an iterative algorithm. Reference [5] employs the Lyapunov optimization to decompose the non-convex problem of minimizing the UAV edge computing system energy consumption into subproblems that optimize the transmission power, RIS phase shift matrix, UAV trajectory, computing resource allocation, and the stability of task queues.

Motivated by these research studies, this paper introduces an effective RIS-aided UAV-MEC scheme for maximizing secure task offloading, where ground user equipment (UEs) offload tasks to a UAV acting as an aerial MEC (AMEC) server equipped with computational resources, while minimiz-ing the total energy consumption of the UAV in the presence of eavesdroppers attempting to capture the confidential tasks being transmitted to the UAV for task offloading. We employ a data-driven algorithm for jointly optimizing the AMEC position, task offloading policy, task offloading scheduling, and RIS phase shifts for optimizing the secrecy energy efficiency (SEE) of these transmissions.

The rest of the paper is organized as follows: Section II formulates the system model and optimization problem. Section III introduces a deep reinforcement learning (DRL) scheme based on the deep deterministic policy gradient (DDPG) for solving the the optimization problem. The numerical analysis of Section IV shows the effectiveness of the proposed approach. Section V draws the conclusions.

## II. System Model

### A. System Model

We consider a scenario where a rotary-wing UAV is deployed as an AMEC server to provide confidential computational offloading and secure content delivery services to $K$ ground UEs distributed within an equal-length $X \times Y$ horizontal grid in the presence of $E$ ground eavesdropping nodes. For a better representation and analysis of the system model dynamics, we consider that the time period $T_a$ is divided into $N$ time slots, where the duration of each time slot is defined as $\delta_{t_s} = \frac{T_a}{N}$, where N is large enough so that channel information remains unchanged within a time slot. The locations of ground UEs and eavesdroppers in time slot $n$ are given by $q_k[n] = [x_k[n], y_k[n]]^T, k \in [1, \ldots, K]$ and $q_E[n] = [x_e[n], y_e[n]]^T, e \in [1, \ldots, E]$, respectively.

The AMEC and all ground UEs are equipped with a single omnidirectional antenna and orthogonal frequency division multiplexing is the employed waveform. Channel access for task offloading transmissions follow the time division multiple access (TDMA) scheme, where only a single user will be selected in each time slot to transmit [6]. The 3-D Cartesian location of the AMEC node in time slot $n$ is defined as $q_A[n] = [x_a[n], y_a[n], z_a[n]]^T$.

The system also features an RIS with coordinates $q_R[n] = [x_r[n], y_r[n], z_r[n]]^T$. The RIS is installed on the surface of one of the surrounding buildings to redirect the signals between AMEC and ground UEs. We assume that the RIS forms a uniform linear array (ULA) of $A$ reflecting elements, as in [7]. The phase shift array in the $n^{th}$ time slot is defined as $\boldsymbol{\phi}[n] \in \mathbb{C}^{O \times O}, \boldsymbol{\phi}[n] = diag\{e^{j\theta_1[n]}, e^{j\theta_2[n]}, \cdots, e^{j\theta_O[n]}\}$, where $\theta_o[n] \in [0, 2\pi), o \in [1, 2, ..., O]$ denotes the phase of the $o^{th}$ element.

Since the ground UEs have limited computing resources, in this considered communication system, each ground UE $k$ offloads a portion of its confidential and time-sensitive task $\Psi_k = \{C_k, D_k, T_k\}$ to the AMEC node through the direct communication path and the indirect transmission path facilitated by the RIS. For task $\Psi_k$, $C_k$ denotes the total number of CPU cycles for task execution, $D_k$ describes the data transferred for task offloading, and $T_k$ is the time taken to complete the computing task. The instantaneous data rate of wireless transmission from user $k$ to the AMEC node for task offloading is expressed as

$$\Omega_k[n] = B_k \log_2 \left(1 + \frac{P_k}{\sigma_k^2} h_k[n]\right), \tag{1}$$

where $B_k$ is the transmission bandwidth available to the $k^{th}$ source UE, $P_k$ is the UE's transmission power, $\sigma_k^2$ is the variance of the additive white Gaussian noise (AWGN) at the UE, and $h_k[n]$ is the channel gain between the AMEC node and the $k^{th}$ UE. Parameter $h_k[n]$ can be expressed under the probabilistic LoS air-to-ground (A2G) communication channel as

$$h_k[n] = \underbrace{\zeta_{LoS} \times g_k^{UAR}[n]}_{\text{LoS UE-RIS-AMEC}} + \underbrace{(1 - \zeta_{LoS}) \times \frac{\gamma_0}{d_{UA}^2[n]}}_{\text{NLoS UE-AMEC}}, \tag{2}$$

where $\zeta_{LoS}$ is the probability of establishing an LoS link, $g_k^{UAR}[n]$ is the UE-RIS-AMEC channel gain, and $\frac{\gamma_0}{d_{UA}^2[n]}$ is the NLoS UE-AMEC channel gain with $\gamma_0$ as the reference path loss and $d_{UA}$ is the distance difference between the ground UE and AMEC. Based on empirical field measurements [8], $\zeta_{LoS}(\theta)$ between the AMEC node and a ground node can be calculated as

$$\begin{aligned} \zeta_{LoS}(\theta) &= \frac{1}{1 + C \exp\left[-B\left(\theta - C\right)\right]}, \\ \theta &= \frac{180}{\pi} \arctan\left(\frac{z_a[n]}{r}\right). \end{aligned} \tag{3}$$

In this context, $C$ and $B$ are constants that depend on the communications environment, such as rural, urban, or densely urban. Parameter $z_a[n]$ represents the height of the AMEC node and $r$ represents the horizontal distance between the AMEC node and the ground receiver. Note that the LoS and NLoS probabilities are related as $\zeta_{NLoS}(\theta) = 1 - \zeta_{LoS}(\theta)$.

The UE-AMEC-RIS channel,

$$g_k^{UAR}[n] = \omega_k^R[n](g_k^{UA}[n] \times \boldsymbol{\phi}[n] \times g_k^{AR}[n]), \tag{4}$$

where $\omega_k^R[n] = \{0, 1\}, \sum_{k=1}^{K} \omega_k^R[n] = 1, \forall n$ denotes whether the RIS will be serving the $k^{th}$ UE or not to ensure only one user will be served in the $n^{th}$ time slot. Parameters $g_k^{UA}[n]$ and $g_k^{AR}[n]$ are the channel gains from the UE to the AMEC node and from the AMEC node to the RIS, respectively. They can be formulated as

$$\begin{aligned} g_k^{UA}[n] &= \frac{\sqrt{\lambda_0}}{d_{UA}^2[n]}\left[1, e^{-j\frac{2\pi}{\lambda}\Upsilon\varphi[n]^{UA}}, \cdots, e^{-j\frac{2\pi}{\lambda}(M-1)\Upsilon\varphi[n]^{UA}}\right], \\ g_k^{AR}[n] &= \frac{\sqrt{\lambda_0}}{d_{AR}^2[n]}\left[1, e^{-j\frac{2\pi}{\lambda}\Upsilon\varphi[n]^{AR}}, \cdots, e^{-j\frac{2\pi}{\lambda}(M-1)\Upsilon\varphi[n]^{AR}}\right], \end{aligned} \tag{5}$$

where $\lambda$ is the carrier wavelength, $\Upsilon$ is the antenna separation, $d_{UA}[n]$ is the distance between the UE and AMEC node, $d_{AR}[n]$ is the distance between the AMEC node and the RIS, and $\varphi[n]^{UA}$ and $\varphi[n]^{AR}$ are the angle of departure (AoD) components of the transmitted signal from the UE to the AMEC node and from the AMEC node to the RIS in the $n^{th}$ time slot, respectively [9].

The confidentiality risk of task offloading under an eavesdropping attack can be quantified using the secrecy capacity. It is defined as the maximum transmission capacity at which an eavesdropper can decode no information, which is calculated as the difference between the capacity of the legitimate channel and that of the wiretap channel [8]:

$$\Omega_{SE}[n] = \max\big((\Omega_k[n] - \max_{e\in E}\Omega_e[n]), 0\big). \tag{6}$$

Parameters $\Omega_k[n]$ and $\Omega_e[n]$ represent the channel capacity of the desired transmission system and the wiretap channel capacity, respectively. The latter can be calculated as

$$\Omega_e[n] = B_e \log_2\left(1 + \frac{P_e\gamma_{e0}}{\sigma_e^2 d_{UE}^2[n]}\right), \tag{7}$$

where $B_e$ is the allocated eavesdropping bandwidth, $P_e$ is the eavesdropping power, $\sigma_e^2$ is the variance of the AWGN at the eavesdropper, $\gamma_{e0}$ is the path loss reference at the eavesdropper, and $d_{UE}^2$ is the distance between the transmitter and the eavesdropper.

The propulsion energy of the rotary-wing UAV as proposed in [10] can be calculated as

$$\varsigma^{A_p}[n] = \delta_{t_s}\left(\iota_c\left(1 + \frac{3\ \nu[n]^2}{\Gamma_{tip}^2}\right) + \frac{1}{2}\ \kappa\ \varepsilon\ \psi\ \Lambda(\nu[n])^3 + \iota_h\left(\sqrt{1 + \frac{\nu[n]^4}{4\nu_o^4}} - \frac{\nu[n]^2}{2\nu_o^2}\right)^{\frac{1}{2}} + \iota_v \nu^v[n]\right), \tag{8}$$

where $\iota_c$ and $\iota_h$ are the constant blade profile power and induced power in hovering status, respectively, $\iota_v$ is the constant power of ascending/descending, $\nu[n]$ and $\nu^v[n]$ are the horizontal and vertical flying speeds of the AMEC node, $\nu_o$ is the mean rotor induced velocity, $\Gamma_{tip}$ is the tip speed of the rotor blade, $\kappa$ is the fuselage drag ratio, $\varepsilon$ is the air density, $\psi$ is the rotor solidity, and $\Lambda \approx \pi r^2$ corresponds to the rotor disc area with rotor radius $r$.

The latency experienced for the offloading of task $\Psi_k$ as shown in [11] can be represented as

$$\mu_k = (1-\alpha_k)\frac{C_k}{c_k^U} + \alpha_k\left(\frac{C_k}{c_k^A} + \frac{D_k}{\Omega_{SE}}\right), \tag{9}$$

where $\alpha_k \in [0,1]$ denotes the proportion of the task offloaded from the $k^{th}$ UE to the AMEC server, whereas $c_k^U$ and $c_k^A$ represent the CPU cycles reserved for the task computation at the UE and AMEC server, respectively. Expression $\Omega_{SE} = \sum_{n=1}^{N}\Omega_{SE}[n]/N$ captures the average secrecy capacity between the UE and the AMEC server. The energy consumption for processing task $\Psi_k$ in part by the AMEC server ($\varsigma_k^{A_c}$) and in part by the UE ($\varsigma_k^{U_c}$) can be formulated as

$$\begin{aligned}\varsigma_k^{A_c} &= \alpha_k G(c_k^A)^{\chi-1}C_k,\\ \varsigma_k^{U_c} &= \alpha_k\frac{D_k}{\Omega_{SE}}P_k + (1-\alpha_k)G(c_k^U)^{\chi-1}C_k,\end{aligned} \tag{10}$$

where $G$ is the effective switched capacitance and $\chi \geq 1$ is a positive constant [11].

## B. Problem Formulation

The objective of this paper is to maximize the total SEE, which is defined as the ratio of the system sum secrecy capacity to its energy consumption [12], for the task offloading channel from the UEs to the AMEC server by maximizing the secrecy capacity while minimizing the total energy consumption of the AMEC node and satisfying the tasks' latency requirements. The optimization problem is defined as

$$\mathrm{P}: \max_{\phi, q_A, \alpha, \omega^R} \frac{\sum_{k=1}^{K}\Omega_{SE}}{\sum_{n=1}^{N}\varsigma^{A_p}[n] + \sum_{k=1}^{K}\varsigma_k^{A_c}}, \tag{11.a}$$

$$s.t.\ \mid e^{j\theta_o} \mid = 1, \forall o, \tag{11.b}$$

$$||q_A[n+1] - q_A[n]|| \leq \{\delta_{t_s}\nu_{max}\}, \tag{11.c}$$

$$\mu_k \leq T_k, \forall k, \tag{11.d}$$

$$0 \leq \alpha \leq 1, \tag{11.e}$$

$$\sum_{k=1}^{K}\alpha_k c_k^U \leq \tau^U, \tag{11.f}$$

$$\omega_k^R[n] = \{0,1\}, \sum_{k=1}^{K}\omega_k^R[n] = 1, \forall n, \forall k. \tag{11.g}$$

Problem $P$ is subject to several constraints: (11.b) is the RIS phase shift constraint, (11.c) is the UAV velocity constraint within its trajectory; (11.d) ensures that task's latency is within its required completion time; (11.e) establishes that fractions of tasks can be offloaded, (11.f) ensures that there are available AMEC resources for task completion, and (11.g) ensures that only one user is being served by the RIS at a time for task offloading transmission.

# III. Proposed Solution

Optimization problem (11) requires a joint approach to AMEC positioning, user selection, task partitioning, and adjustment of the RIS reflecting elements in the presence of eavesdropping. The objective function of (11) is non-concave because of the correlations among the optimization variables along with the unit modulus constraint of $\phi$ [7]. Therefore, the optimization problem is found to be a non-convex and non-trivial optimization problem [13]. Since there is no standard method for solving such a non-convex optimization problem, we investigate solving the problem employing data-driven solutions instead of conventional mathematical optimization tools. Most of the traditional solutions to equivalent multi-parameter optimization problems are iterative and alternately optimize the parameters reaching suboptimal results [12]. Alternatively, we propose a DDPG algorithm for establishing a continuous action space [14] and solve the non-convex problem of this paper. The solution maximizes the SEE while minimizing the energy consumption of the AMEC server and satisfying the latency requirements of the offloaded user tasks.

## A. MDP Formulation

The AMEC server acts as a DDPG agent with a Markov decision process (MDP) composed of the state space $\mathcal{S}$,

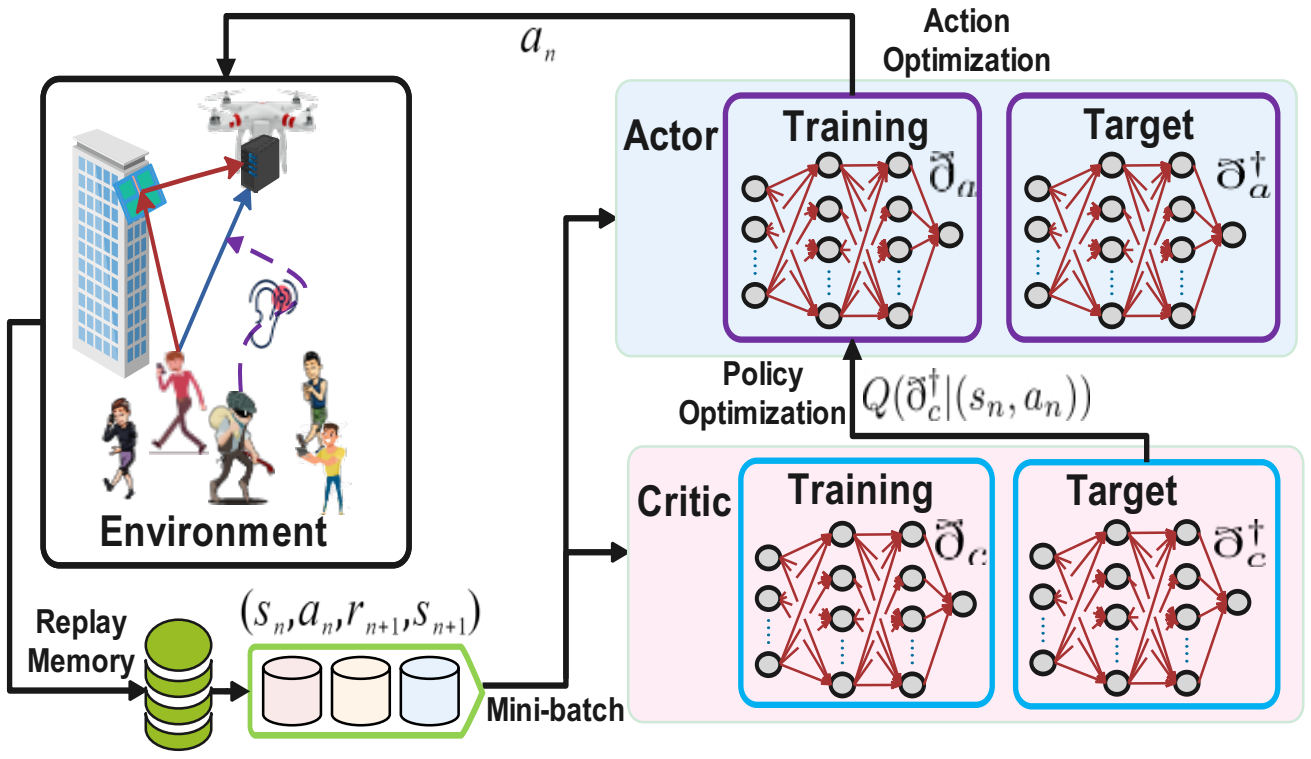

Fig. 1: Block diagram of the proposed DDPG architecture.

the action space $\mathcal{A}$, the reward space $\mathcal{R}$, and the transition probability space $\mathcal{T}$: $(\mathcal{S}, \mathcal{A}, \mathcal{R}, \mathcal{T})$.

**State:** The set of states is defined as $\mathcal{S} = \{s_1, ..., s_n, .., s_N\}$. State $s_n = \{s_A^Q, s_A^V, s_A^E, s_A^C, s_R^\phi, s_U^S, s_U^T, s_U^C\}$ captures all relevant system features: AMEC node position $s_A^Q$, AMEC node velocity $s_A^V$, AMEC node energy $s_A^E$, available computing resources $s_A^C$ for task offloading, RIS phase shifts $s_R^\phi$, task size $s_U^S$ at the UE, task completion time $s_U^T$ at the UE, and available computing resources $s_U^C$ at the UE.

**Action:** The states are transited according to the defined actions. A set of actions is defined as $\mathcal{A} = \{a_1, ..., a_n, .., a_N\}$. Action where each action at time $a_n = \{a_A^V, a_A^Q, a_R^\phi, a_U^\alpha, a_U^\omega\}$ comprises the UAV velocity $a_A^V$, the UAV flight direction $a_A^Q$, RIS phase shift adjustments $a_R^\phi$, task offloading decision $a_U^\alpha$, and UE selection $a_U^\omega$. Actions $a_U^\alpha$ and $a_U^\omega$ are introduced as continuous probability variables to deal with the binary variables $\alpha$ and $\omega_k^R$. If $a_U^\alpha$ or $a_U^\omega \geq 0.5$, then $\alpha$ or $\omega_k^R$ are rounded up to 1 or else to 0 [15].

**Reward:** After taking action $a_n$ in state $s_n$ at time slot $n$, the agent will receive reward $r_n(s_n, a_n)$. The reward function, $r_n(s_n, a_n) = \frac{\sum_{k=1}^{K} \Omega_{SE}[n]}{\varsigma^{A_p}[n] + \sum_{k=1}^{K} \varsigma_k^{A_c}}$, is designed to support the desired SEE performance objective.

## B. Deep Deterministic Policy Gradient

The DDPG algorithm leverages the actor and the critic networks as depicted in Fig. 1. The actor network is employed to formulate a deterministic policy that optimizes the output of the critic network. It processes the state inputs and produces deterministic actions. The critic network is designed to approximate the Q-value function, thereby assessing the deterministic policy generated by the actor network. It receives the deterministic policy from the actor network as its input and outputs the corresponding Q-values. Both the actor and the critic network consist of a training and a target network. Consequently, the DDPG algorithm encompasses a total of four neural networks [16].

When the agent takes an action, the system generates an experience record. At time step $n$, this experience comprises the current state $s_n$, the action $a_n$, the reward $r_n$, and the subsequent state $s_{n+1}$, forming tuple $e_n = (s_n, a_n, r_n, s_{n+1})$. Each such experience is stored in a replay memory with a capacity of $\aleph$, resulting in the memory set $\mathcal{M} = \{e_1, ..., e_t, ..., e_\aleph\}$. The DRL model is then updated by sampling mini-batches from this replay memory, rather than relying solely on the most recent state transition.

The updates of the training critic network are obtained as

$$\eth_c = \eth_c - \varrho_c \, \Delta_{\eth_c} \ell(\eth_c), \tag{12}$$

where $\eth_c$ represents the weights and the bias of the training critic network, $\varrho_c$ is the learning rate, and $\Delta_{\eth_c^t}$ is the gradient. Parameter $\ell(\eth_c)$ is the loss function of the training critic network, which can be computed as

$$\ell(\eth_c) = \mathbb{E}\Bigg[\Bigg(\Big[r_n + \zeta \times \; Q\Big(\eth_c^\dagger \mid (s_{n+1}, \tilde{a})\Big)\Big] - \Big[Q\Big(\eth_c \mid (s_n, a_n)\Big)\Big]\Bigg)^2\Bigg], \tag{13}$$

where $\tilde{a}$ is the action of the agent that follows the deterministic policy drafted by the target actor network whereas $\eth_c^\dagger$ captures the network's weights and bias. The training network updates occur more frequently than the target network updates.

The parameters updates for the training actor network are obtained as

$$\eth_a = \eth_a - \varrho_a \, \Delta_a Q\Big(\eth_c^\dagger \mid (s_n, a_n)\Big) \, \Delta_{\eth_a} \mho(\eth_a \mid s_n), \tag{14}$$

where $\eth_a$ corresponds to the weights and bias of the training actor network $\mho(\eth_a \mid s_n)$, $\varrho_a$ is the learning rate, $\Delta_a Q(\cdot)$ is the gradient of the target critic network output with respect to the action taken, and $\Delta_{\eth_a} \mho(\cdot)$ is the gradient of the training actor network given $\eth_a$. The gradient $\Delta_a Q(\cdot)$ is incorporated into the update of the training actor network to ensure that the subsequent action selection by the critic network optimizes the Q-value function.

The target critic and target actor network updates are obtained as

$$\eth_c^\dagger \leftarrow \eta_c \, \eth_c + (1 - \eta_c) \, \eth_c, \tag{16}$$

$$\eth_a^\dagger \leftarrow \eta_a \, \eth_a + (1 - \eta_a) \, \eth_a, \tag{15}$$

where $\eta_c$ and $\eta_a$ are the learning rates for updating the critic and actor networks, respectively.

# IV. Simulation Results

We numerically analyze the performance of the proposed scheme for optimizing the SEE of the RIS-assisted AMEC system while minimizing the total energy consumption of the AMEC server and satisfying the task latency requirements. The environment consists of 6 ground UEs and 3 eavesdroppers that are randomly distributed in a $(600, 400)$ $m$ square area. The RIS is deployed on the surface of one of the surrounding buildings at $(200, 200, 20)$ with $O = 64$ reflection elements. The AMEC server initiates its mission with a 140 $KJ$ energy level and the AMEC mission time is set to $N = 200$ time slots. Table I provides the parameters of the simulated scenarios and the hyperparameters of the proposed DDPG algorithm. Both the critic and actor networks of the proposed DDPG algorithm consist of the input layer, output layer, and two fully connected hidden layers with 80 and 40 neurons, respectively. The state space size is $I_s' = 2A + 8$, which is equivalent to the number of the neurons of the input layer, where $2A$ captures the changes of each complex beamforming vector of the $A$ reflecting elements and 8 represents the number of other parameters stated in the previous section for each state. Expression $O_a' = 2A + 4$ captures the size of the action space which corresponds to the number of neurons of the output layer.

TABLE I: Simulation Parameters.

| Parameters | Values | Parameters | Values |
|---|---|---|---|
| $\Gamma_{tip}, v_o, v_{max}$ | $120, 4, 10\ m/s$ | $\aleph$ | $10^6$ |
| $\kappa, \varepsilon, \psi, \Lambda$ | 0.6,0.051.23,0.5 | $\varrho_c = \varrho_a$ | $10^{-3}$ |
| $\iota_c, \iota_h, \iota_v$ | 79.4, 89.4,11.46 $W$ | $\eta_c = \eta_a$ | $10^{-3}$ |
| $\sigma_k^2, \sigma_e^2$ | $10^{-12}$ | $\delta_c = \delta_a$ | $10^{-5}$ |
| $C, B$ | 9.61, 0.16 | *DDPG episodes* | $12 * 10^4$ |
| $\lambda_0$ | $-70\ dB$ | *Batch size* | 64 |
| $B_k$ | $1\ MHz$ | *DDPG time slots* | 750 |
| $f_c$ | $3.2\ GHz$ | $T_k, D_k$ | $30\ s, 512 \sim 1024\ Kb$ |
| $P_k, P_e$ | $1\ mW$ | $C_k, \tau^U$ | $10^3$ Cycles/bit |

Fig. 2 illustrates the efficacy of the proposed DDPG-based algorithm for optimizing the SEE of the system. It plots the SEE, which corresponds to the reward function that we defined earlier, over the learning episodes for the proposed DDPG solution with 64 and 32 RIS elements and a deep Q-learning (DQL)-based technique using the same objective and reward functions. The DQL-based approach utilizes a similar DNN architecture as employed in both the critic and actor networks of the DDPG. It can be observed that the DDPG-based solution outperforms the DQL baseline. This is so because of the DDPG's capability to produce a specific action due to its deterministic policy nature rather than a value estimation of the DQL for action selection as an off-policy value-based method. On the other hand, the DQL approach converges faster as the DQL excels in discrete action spaces with a simpler implementation. We further observe that as the number of reflecting elements increases, the SEE performance improves. This enhancement is attributed to the greater degrees of freedom for signal reflection optimization, leading to better signal alignment at the receiver yielding to improving the SNR and the overall communication quality. However, as the number of RIS-reflecting elements increases, the computational complexity of the solution increases as well, resulting in a longer convergence time of the proposed solution because of the larger action and observation spaces that the DDPG algorithm must manage and learn from.

Figure 3 illustrates the optimized 3D trajectory obtained for the UAV utilizing the proposed DDPG-based strategy. The UAV takes off at $(0, 0)$ and the trajectory optimization is set to begin after reaching an altitude of 60 $m$. After that, the UAV carrying the AMEC server strategically navigates its path between different ground users based on the DDPG algorithm while maneuvering the environment blockage (buildings) to boost the LoS communications probability with the help of the RIS to enhance the signal quality given the ground user locations/channels while minimizing the exposure to eavesdroppers. The result illustrates the robustness of the DDPG algorithm in handling uncertainties within the network environment via the dynamic adaptability of the optimized trajectory in response to varying ground users' locations and computation offloading requirements over time.

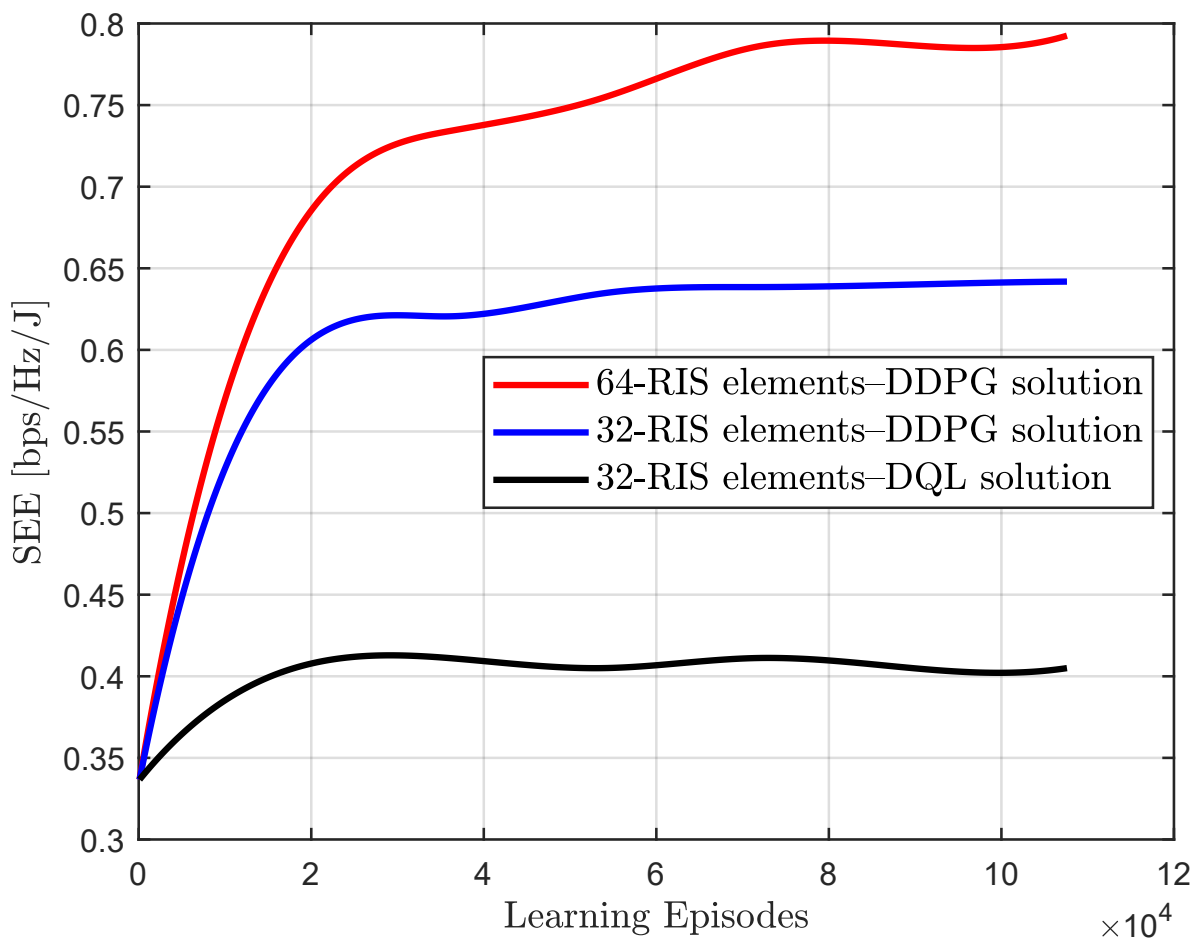

Fig. 2: SEE performance for different numbers of RIS-reflecting elements versus learning episodes for the proposed DDPG solution and DQL benchmark technique.

Figure 4 presents the AMEC energy level over its flight time represented in time slots. The goal is assessing the proposed algorithm's performance and examine the importance of the individual components in minimizing the AMEC energy consumption. The figure highlights the algorithm's success in preserving approximately 65% energy of the initial AMEC energy by the end of its operation. Without RIS phase shift optimization, only 55% energy capacity remains at the end of the mission, underscoring the RIS phase shift optimization's pivotal role in enhancing the energy minimizing of the AMEC operation. Without UAV trajectory optimization but with RIS phase shift optimization, the remaining energy level drops to 50% of its initial capacity, pointing to the critical importance of optimizing the UAV trajectory for conserving energy.

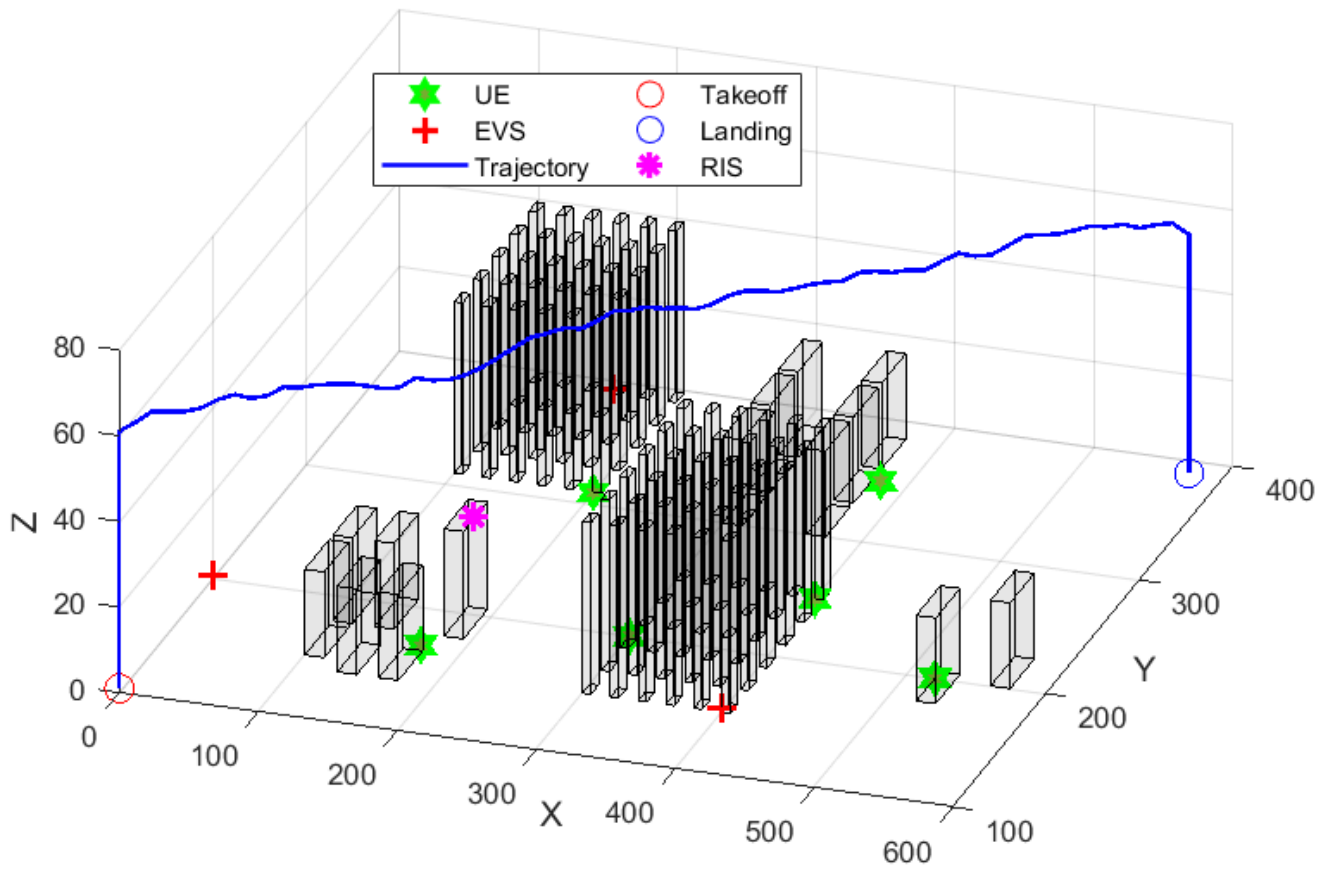

Fig. 3: DDPG-based AMEC optimized trajectory.

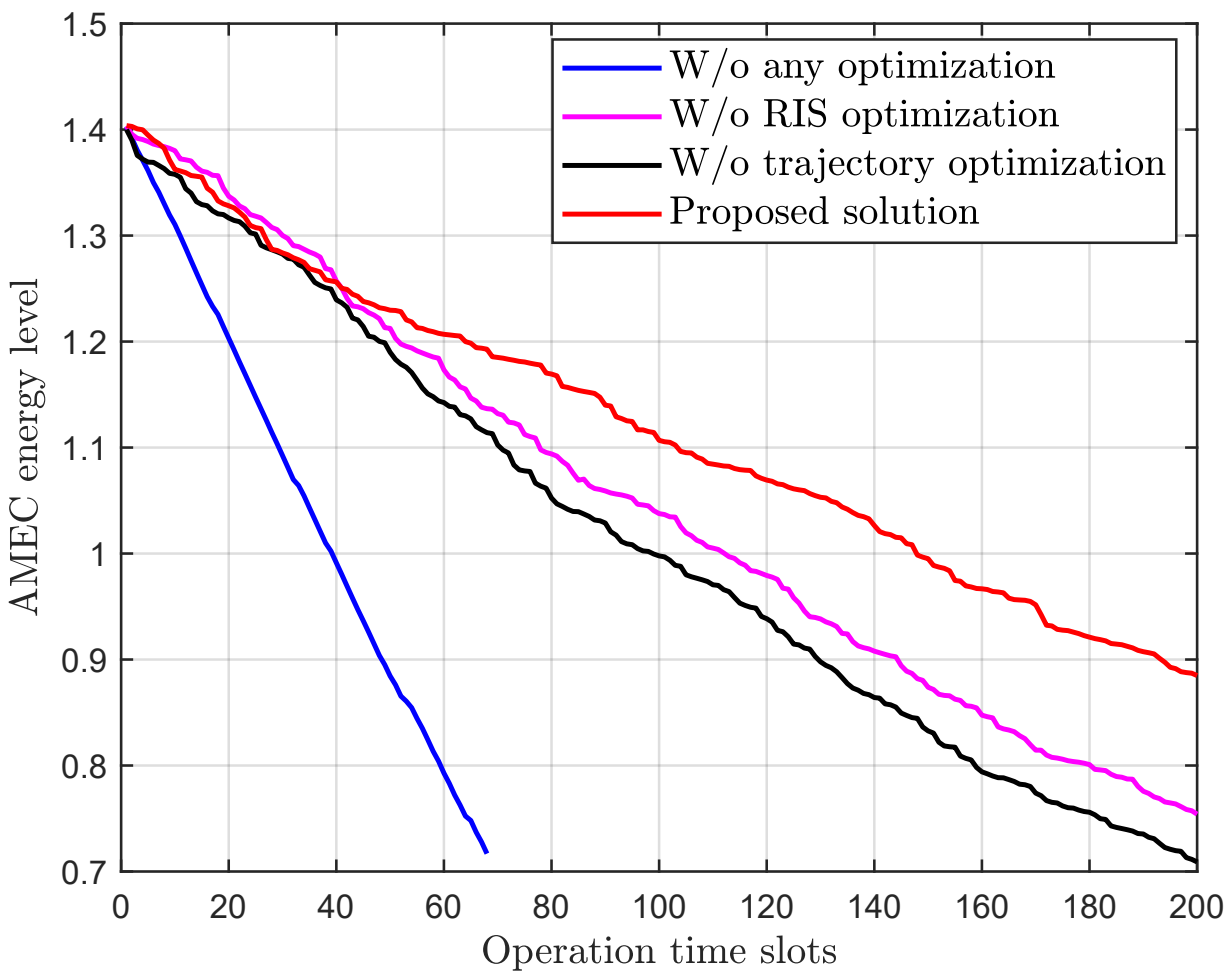

Fig. 4: AMEC energy level over operation time slots for the proposed solution and other techniques.

## V. Conclusions

In this paper, we have investigated the potential of RIS-aided AMEC for securing the legitimate task offloading transmission of multiple ground users to the aerial edge server in the presence of eavesdropping attacks. The UAV trajectory, ground user selection, task offloading proportion, and the RIS phase shifts need to be jointly optimized to maximize the SEE while minimizing the total energy consumption of the AMEC server and satisfying task completion latency requirements. We have proposed a DDPG-based algorithm to determine the control policy for the continuous UAV path planning, users selection, task offloading, and RIS phase shifter action spaces of the system. Numerical results have shown the capability of the proposed solution to maximize the SEE performance while preserving the energy level of the AMEC server and satisfying the QoS requirements.

## Acknowledgement

This work was supported in part by the NSF PAWR program, under grant number CNS-1939334 and by the Office of Naval Research under Award No. N00014-23-1-2808.